\documentclass[aps,pra,reprint,longbibliography,dvipdfmx]{revtex4-2}

\usepackage{amsmath,amssymb,bm}
\usepackage{graphicx}
\usepackage{xcolor}
\usepackage{multirow}
\usepackage{tabularx}
\usepackage{physics}
\usepackage{mathrsfs}

\begin{document}

\title{Structured-light-mediated hybrid entanglement between photon polarization and electronic orbital angular momentum}

\author{Hiroaki Saito}
\affiliation{Department of Physics and Electronics, Osaka Metropolitan University, 1-1 Naka-ku, Sakai, Osaka 599-8531, Japan}
\author{Nobuhiko Yokoshi}
\affiliation{Department of Physics and Electronics, Osaka Metropolitan University, 1-1 Naka-ku, Sakai, Osaka 599-8531, Japan}

\date{\today}

\begin{abstract}
We propose a minimal quantum-optical scheme for generating hybrid entanglement between photon polarization and electronic orbital angular momentum in a semiconductor quantum disk. A spin--orbit structured two-photon state excites two channels in the same disk: a radiatively recombining zero-orbital-angular-momentum channel and a finite-orbital-angular-momentum channel that stores the electronic orbital qubit. An effective coherent mapping prepares a selected two-excitation state, followed by emission of a photon whose polarization is entangled with the residual electronic orbital state. A master-equation analysis shows that the heralded state conditioned on single-photon occupation of the selected output mode approaches the target entangled state in the ideal coherent limit. We also discuss orbital relaxation and perturbative validity conditions for branch-dependent Coulomb shifts, orbital-angular-momentum mixing, and finite-orbital-angular-momentum radiative leakage. This proof-of-principle effective model suggests a route toward structured-light-mediated photon--electron hybrid entanglement in semiconductor nanostructures.
\end{abstract}

\maketitle

\section{Introduction}

The coherent control of quantum states in semiconductor nanostructures is a key ingredient for quantum information processing. Electron spins in semiconductors can be manipulated electrically~\cite{noiri2022fast} and optically with pulsed light~\cite{liu2010quantum}. In particular, coherent transfer of photon polarization to electron spin has been demonstrated in V-type three-level semiconductor structures~\cite{kosaka2008coherent,vrijen2001spin,yokoshi2013creation}. Structured optical fields provide an additional route to controlling electronic states through the spatial and angular-momentum structure of light. Laguerre--Gaussian and related vector-vortex beams carry orbital angular momentum (OAM) and exhibit structured phase and polarization profiles~\cite{allen1992orbital}. Recent experiments have shown that such optical structures can be transferred or imprinted onto electronic spin and orbital degrees of freedom in semiconductors~\cite{matsumoto2024coherent,ishihara2023imprinting}. These developments motivate the use of optical spin and OAM selection rules as building blocks for coherent light--matter mappings.

Electronic orbital states are natural candidates for coupling semiconductor nanostructures to structured light, because their spatial wave functions can reflect the angular-momentum content of the optical field. Under suitable device geometries, orbital states may also be converted into charge-sensitive signals through tunnel coupling, charge response, or dispersive readout~\cite{vandersypen2017interfacing,schleser2004time,petersson2012circuit}. Since electronic OAM states are not restricted in principle to a two-level subspace, they may further provide a route to higher-dimensional electronic encodings. In this work, we focus on the simplest case: an electronic OAM qubit formed by two finite-OAM orbital states.

Entanglement between photonic and material degrees of freedom is central to quantum communication and quantum networks~\cite{tanzilli2005photonic}. Hybrid entanglement involving different optical degrees of freedom has been explored in photonic systems~\cite{morin2014remote,jeong2014generation}, while photon--spin entanglement has been demonstrated in semiconductor quantum-dot platforms~\cite{degreve2012quantum,gao2012observation,schaibley2013demonstration}. A natural extension is hybrid entanglement between the polarization of a photon and the OAM of a stationary electronic excitation. Such an interface would connect photonic polarization qubits to electronic orbital states in semiconductor nanostructures.

Spin--orbit structured optical fields are well suited for this purpose because they encode polarization and OAM within a single photonic state. Vector-vortex states, higher-order Poincar\'e-sphere states, and related optical fields provide representative examples~\cite{mclaren2015measuring,krenn2017orbital,yi2014hybrid,fickler2014quantum,d2016entangled}. More general polarization textures, including optical skyrmions, further illustrate the controllable combination of spin and orbital angular momentum in structured light~\cite{teng2023physical,sugic2021particle}. In semiconductor nanostructures, this spin--orbit structure can be combined with optical selection rules to map photonic polarization and OAM components onto electronic degrees of freedom.

Here we propose a minimal quantum-optical scheme for generating hybrid entanglement between photon polarization and electronic OAM in a semiconductor nanostructure. The scheme uses a spin--orbit structured two-photon state and two optically addressed excitation channels in the same disk. One channel has zero OAM and can recombine radiatively, while the other has finite OAM and stores the residual electronic orbital qubit. Radiative recombination of the zero-OAM excitation emits a photon whose polarization becomes entangled with the remaining finite-OAM electronic state.

The aim of this work is to demonstrate the basic principle of photon--electron OAM hybrid-entanglement generation within a minimal effective model, rather than to provide a device-specific microscopic theory. We formulate the process as an effective coherent mapping consistent with angular-momentum conservation and optical selection rules, and analyze the subsequent emission dynamics using a master-equation treatment. We also include orbital relaxation as a representative decoherence channel and discuss perturbative validity conditions for branch-dependent Coulomb shifts, OAM mixing, and finite-OAM radiative leakage. Within this proof-of-principle framework, the proposed scheme suggests a route toward structured-light-mediated hybrid quantum interfaces in semiconductor nanostructures.

\section{Physical background and framework}
\label{sec:physical_background}

We define the optical input state and the effective light--matter mappings used in the model. Throughout this work, we use a parallel notation for photonic and electronic degrees of freedom. A photon state is written as $\ket{\mu,l}_{\eta}$, where $\mu=L,R,H,V$ denotes polarization, $l$ denotes optical OAM, and $\eta=A,B,\gamma$ labels the optical component or emitted mode. An electronic excitation state is written as $\ket{\sigma,m}_{\eta}$, where $\sigma=\uparrow,\downarrow,+$ denotes electron spin, $m$ denotes electronic OAM, and $\eta=A,B$ labels the excitation channel. Thus, optical polarization is mapped to electron spin, while optical OAM is mapped to electronic OAM.

We use the circular polarization basis $\{\ket{L},\ket{R}\}$ and the linear polarization basis $\{\ket{H},\ket{V}\}$, with
\begin{align}
    \ket{H}_p = \frac{1}{\sqrt{2}}(\ket{L}_p+\ket{R}_p),\quad
    \ket{V}_p = \frac{1}{\sqrt{2}}(\ket{L}_p-\ket{R}_p).
    \label{eq:linear_circular_convention}
\end{align}
Here the subscript $p$ indicates that Eq.~\eqref{eq:linear_circular_convention} acts only on the polarization degree of freedom; the OAM label is suppressed. When the OAM label is included, the same convention is understood as $\ket{H,l}_{\eta}=(\ket{L,l}_{\eta}+\ket{R,l}_{\eta})/\sqrt{2}$ and $\ket{V,l}_{\eta}=(\ket{L,l}_{\eta}-\ket{R,l}_{\eta})/\sqrt{2}$. This equation fixes the relative phase convention used throughout the paper.

We consider a spin--orbit structured two-photon input state of the form
\begin{align}
\ket{\psi_{\rm in}}
=
\frac{1}{\sqrt{2}}
\left(
\ket{L,0}_{A}\ket{H,+2}_{B}
+
\ket{R,0}_{A}\ket{H,-2}_{B}
\right).
\label{eq:input_photon_state}
\end{align}
Such correlated polarization--OAM photon-pair states can be prepared from polarization-entangled photon pairs using wave plates, spin--orbit conversion elements, and polarization optics, as demonstrated in spontaneous-parametric-down-conversion, entangled vector-vortex, and related structured-light experiments~\cite{kwiat1995new,marrucci2006optical,Forbes:16,fickler2014quantum,d2016entangled}. Here $A$ and $B$ label two optical components that address two excitation channels in the same semiconductor quantum disk. The $A$ component carries zero OAM and creates the radiatively recombining excitation channel, while the $B$ component carries finite OAM and creates the residual electronic OAM channel. The detailed optical preparation circuit is not essential for the present model; the relevant point is that the relative amplitudes and phases in Eq.~\eqref{eq:input_photon_state} can be coherently controlled.

We next specify the effective optical transfer to semiconductor electronic states. In a GaAs-like V-type three-level structure, right- and left-circularly polarized light selectively address different electron spin states from a Zeeman-split light-hole valence-band state~\cite{vrijen2001spin,kosaka2008coherent,yokoshi2013creation}. The relevant light-hole states may be written as $\ket{\Psi^\pm}=(\ket{m_j=1/2}\pm\ket{m_j=-1/2})/\sqrt{2}$, where $\ket{m_j}$ denotes a total-angular-momentum eigenstate of the valence band. We adopt the convention $\ket{L}_p\rightarrow\ket{\downarrow}_s$ and $\ket{R}_p\rightarrow\ket{\uparrow}_s$. More generally, the coherent polarization-to-spin transfer demonstrated in semiconductor V-type systems realizes the qubit mapping
\begin{align}
    \alpha\ket{L}_p+\beta\ket{R}_p
    \rightarrow
    \alpha\ket{\downarrow}_s+\beta\ket{\uparrow}_s,
    \quad
    |\alpha|^2+|\beta|^2=1,
    \label{eq:general_polarization_spin_mapping}
\end{align}
up to a fixed and calibratable phase convention. With Eq.~\eqref{eq:linear_circular_convention}, this gives
\begin{align}
    \ket{H}_p
    \rightarrow
    \ket{+}_s
    =
    \frac{1}{\sqrt{2}}
    \left(
    \ket{\downarrow}_s+\ket{\uparrow}_s
    \right).
    \label{eq:H_polarization_to_spin_plus}
\end{align}
Equations~\eqref{eq:general_polarization_spin_mapping} and~\eqref{eq:H_polarization_to_spin_plus} describe the polarization-to-spin part of the mapping; OAM labels are suppressed there and restored in the channel-specific mappings below. Equation~\eqref{eq:H_polarization_to_spin_plus} is a special case of the established polarization-qubit transfer, not an additional assumption. The reverse process is used during radiative recombination: within the same convention, $\ket{\downarrow}_s$ emits a left-circularly polarized photon and $\ket{\uparrow}_s$ emits a right-circularly polarized photon.

\begin{figure}[t]
\includegraphics[width=\linewidth]{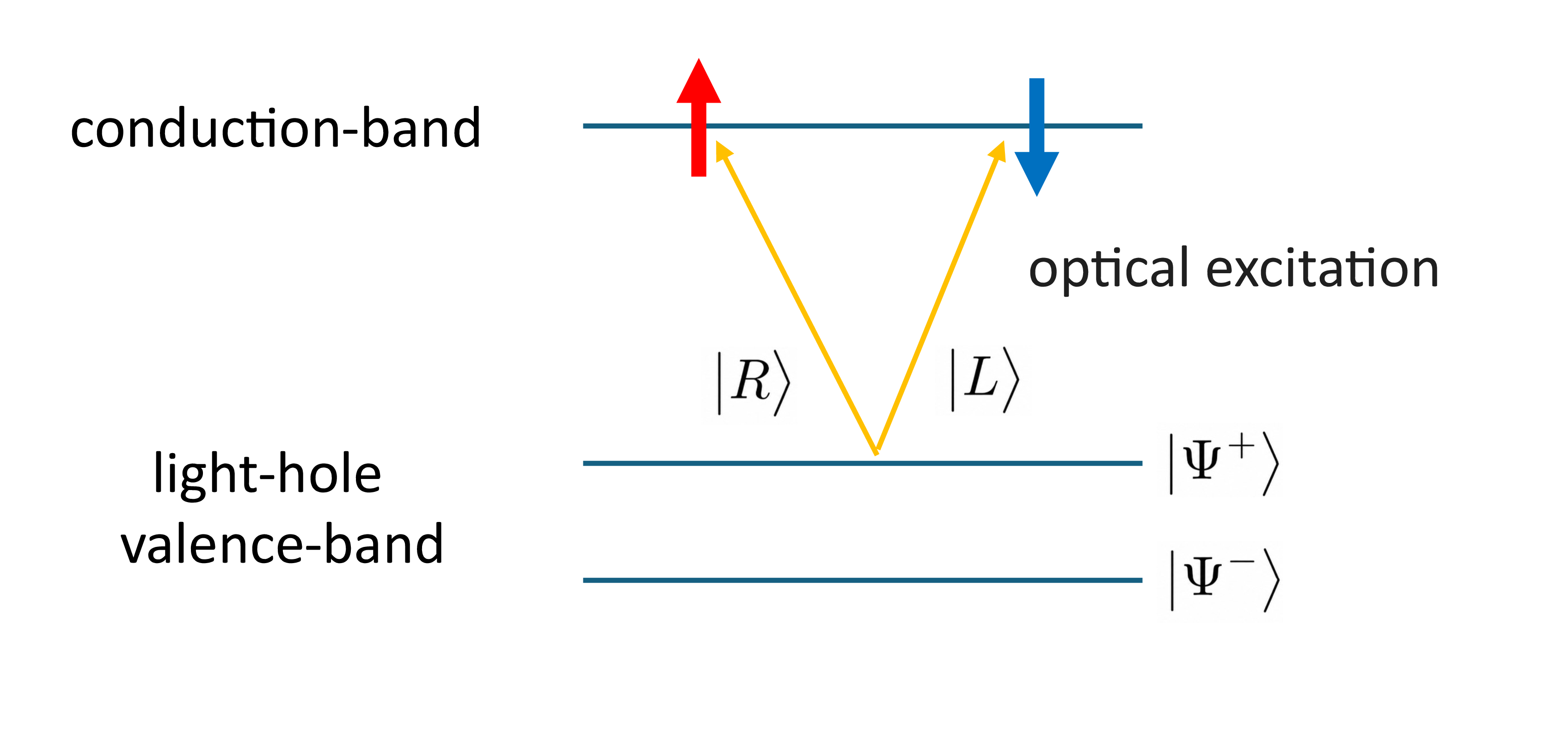}
\caption{Spin-excitation selection rules in a GaAs-based V-type three-level semiconductor structure. Left- and right-circularly polarized photons selectively excite different electron spin states from one of the Zeeman-separated light-hole states, $\ket{\Psi^-}$ or $\ket{\Psi^+}$.}
\label{fig:v_type_selection_rules}
\end{figure}

Figure~\ref{fig:v_type_selection_rules} schematically illustrates the spin-selective optical transitions. The finite-OAM channel is motivated by experiments on structured-light control in semiconductors. Higher-order polarization states of light have been coherently transferred to electronic spin-structure states~\cite{matsumoto2024coherent}, indicating that spatially structured optical superpositions can preserve relative phase information during transfer. Optical initialization and manipulation of higher-order electronic states carrying spin and OAM have also been demonstrated in a semiconductor quantum disk~\cite{terashima2025optical}. Related experiments on excitonic OAM transfer by optical vortex fields establish coupling between optical OAM and excitonic orbital degrees of freedom~\cite{ueno2009coherent,shigematsu2016coherent}, and vector-vortex beams have been used to imprint spatial helicity structures onto spin textures in semiconductors~\cite{ishihara2023imprinting}. These results support the use of structured light as a coherent control resource for finite-OAM electronic states.

For the finite-OAM channel $B$, we adopt the effective single-excitation mapping
\begin{align}
    \ket{H,\pm2}_{B}
    \rightarrow
    \ket{+,\pm2}_{B},
    \label{eq:finite_oam_effective_mapping}
\end{align}
where the optical OAM index selects the corresponding electronic OAM state and the linear polarization component is transferred to the electron spin state $\ket{+}_s$. Here $\ket{+,\pm2}_{B}\equiv\ket{+}_{s,B}\otimes\ket{\pm2}_{o,B}$, so that the notation is parallel to the photonic state $\ket{H,\pm2}_{B}$. Equation~\eqref{eq:finite_oam_effective_mapping} is not intended as a microscopic derivation of the semiconductor optical response. Rather, it is the effective representation used here for the experimentally supported coherent transfer of higher-order optical structure to electronic spin--OAM states, based on Refs.~\cite{matsumoto2024coherent,terashima2025optical}. The essential assumption is that the two addressed branches, $\ket{H,+2}_{B}$ and $\ket{H,-2}_{B}$, are transferred to the corresponding electronic OAM states while preserving their relative amplitudes and phase.

The logical status of the mappings used below is as follows. The $A$-channel polarization-to-spin transfer is an experimentally established coherent qubit interface. The $B$-channel finite-OAM transfer is an experimentally supported effective mapping based on coherent higher-order optical transfer and optical control of spin--OAM electronic states in quantum disks. The new element of this work is to combine these single-excitation mappings branch by branch for the spin--orbit structured two-photon input in Eq.~\eqref{eq:input_photon_state}, and then to use radiative recombination of the zero-OAM excitation to generate photon--electron OAM hybrid entanglement.

Accordingly, the state mappings introduced in Sec.~\ref{sec:model} should be read as effective coherent mappings obtained by applying the above single-excitation processes to each branch of the entangled two-photon input. By linearity, the relative amplitudes and phases of Eq.~\eqref{eq:input_photon_state} are transferred to the corresponding two-excitation state, provided that the absorption process does not generate which-path information and that branch-dependent energy shifts are negligible during the heralding time. These requirements are not assumed to hold automatically; Sec.~\ref{sec:model} discusses branch-dependent Coulomb shifts, OAM mixing, and finite-OAM radiative leakage.

\section{Model}
\label{sec:model}

We now introduce a minimal effective model for generating hybrid entanglement between photon polarization and electronic orbital angular momentum. The model is intended to describe the principle of the entanglement-generation process, rather than a microscopic theory of a specific semiconductor device. We use the coherent polarization-to-spin and structured-light-to-orbital transfer processes discussed in Sec.~\ref{sec:physical_background} as effective building blocks~\cite{kosaka2008coherent,matsumoto2024coherent,terashima2025optical}.

The labels $A$ and $B$ denote two optically addressed excitation channels within a single semiconductor quantum disk. Channel $A$ is a zero-OAM excitation channel that recombines radiatively into the selected optical mode, whereas channel $B$ is a finite-OAM electronic orbital channel that remains as the stationary OAM qubit. We use the term ``excitation'' to denote the optically generated electronic degree of freedom together with the corresponding valence-band configuration. After radiative recombination in channel $A$, the relevant stationary degree of freedom is the finite-OAM electronic state in channel $B$. In this effective description, the hole degree of freedom in channel $B$ is assumed to be inert, and we focus on the electronic orbital component of the residual excitation.

The effective light--matter interaction responsible for absorption is written in rotating-wave form as
\begin{align}
H_{\rm int}(t)
=
\hbar
\sum_{\mu,l,\eta}
\Omega_{\mu l}^{(\eta)}(t)
X^{\dagger}_{\mu,l,\eta}
a_{\mu,l,\eta}
+
{\rm H.c.},
\label{eq:effective_light_matter_hamiltonian}
\end{align}
where $a_{\mu,l,\eta}$ annihilates a photon in optical component $\eta=A,B$ with polarization $\mu$ and OAM $l$, and $X^{\dagger}_{\mu,l,\eta}$ creates the corresponding optically allowed electronic excitation. The matrix element $\Omega_{\mu l}^{(\eta)}(t)$ includes the spatial overlap between the optical mode and the electronic envelope functions. Equation~\eqref{eq:effective_light_matter_hamiltonian} is a compact effective representation of the spin and OAM selection rules used below.

The state $\ket{G}$ denotes the zero-excitation state of the full system. For the zero-OAM channel $A$, the polarization selection rule gives
\begin{align}
X^{\dagger}_{L,0,A}\ket{G}
=
\ket{\downarrow,0}_{A},
\quad
X^{\dagger}_{R,0,A}\ket{G}
=
\ket{\uparrow,0}_{A}.
\label{eq:A_channel_creation}
\end{align}
For the finite-OAM channel $B$, the structured-light coupling gives
\begin{align}
X^{\dagger}_{H,\pm 2,B}\ket{G}
=
\ket{+,\pm 2}_{B}.
\label{eq:B_channel_creation}
\end{align}
Equation~\eqref{eq:B_channel_creation} assumes that the optical OAM component with $l=\pm2$ addresses the corresponding finite-OAM electronic orbital state while preserving the spin superposition associated with the incident linear polarization. The phase convention for $\ket{+}_s$ is the same as that in Eq.~\eqref{eq:H_polarization_to_spin_plus}.

Applying these selection rules branch by branch to the input state in Eq.~\eqref{eq:input_photon_state} gives
\begin{align}
&\frac{1}{\sqrt{2}}
\left(
\ket{L,0}_{A}\ket{H,+2}_{B}
+
\ket{R,0}_{A}\ket{H,-2}_{B}
\right)
\nonumber\\
&\quad \rightarrow
\frac{1}{\sqrt{2}}
\left(
\ket{\downarrow,0}_{A}\ket{+,+2}_{B}
+
\ket{\uparrow,0}_{A}\ket{+,-2}_{B}
\right)
\equiv
\ket{\psi_{\rm ex}} .
\label{eq:two_excitation_mapping}
\end{align}
Equivalently,
\begin{align}
  \ket{\psi_{\rm ex}}
=
\frac{1}{\sqrt{2}}
\left(
\ket{\downarrow,0}_{A}\ket{+,+2}_{B}
+
\ket{\uparrow,0}_{A}\ket{+,-2}_{B}
\right).
\label{eq:biexciton_state}
\end{align}
Equation~\eqref{eq:two_excitation_mapping} is the effective-state representation of Eq.~\eqref{eq:effective_light_matter_hamiltonian} together with the selection rules in Eqs.~\eqref{eq:A_channel_creation} and \eqref{eq:B_channel_creation}. Its validity requires that the two absorption branches remain indistinguishable except for the encoded spin--OAM quantum numbers, so that no which-path information is generated during excitation.

We make two additional assumptions. First, zero-OAM excitations recombine radiatively much more efficiently than finite-OAM excitations because they have a larger overlap with the selected optical emission mode. Thus, on the emission timescale considered here, the $m=0$ excitation in channel $A$ recombines and emits a photon, while the finite-OAM electronic state in channel $B$ remains as the stationary degree of freedom. The recombination rule may be written as
\begin{align}
\ket{\downarrow,0}_{A}\ket{+,m}_{B}
&\rightarrow
\ket{L,0}_{\gamma}\ket{+,m}_{B},
\nonumber\\
\ket{\uparrow,0}_{A}\ket{+,m}_{B}
&\rightarrow
\ket{R,0}_{\gamma}\ket{+,m}_{B},
\quad
m=\pm2 ,
\label{eq:recombination_example}
\end{align}
where $\gamma$ denotes the emitted photon mode. In the dynamical calculation below, we include orbital relaxation from the finite-OAM states to the zero-OAM state as a representative mechanism that erases the stored OAM information.

Second, the relative amplitudes and phases in Eq.~\eqref{eq:two_excitation_mapping} are assumed to be preserved during absorption. This assumption is motivated by phase-coherent polarization-to-spin transfer in semiconductor systems~\cite{kosaka2008coherent}, coherent transfer of higher-order optical polarization states to electronic spin-structure states~\cite{matsumoto2024coherent}, and optical control of higher-order spin--OAM electronic states in semiconductor quantum disks~\cite{terashima2025optical}. Coulomb-induced correlations, biexciton binding effects, and the probabilistic nature of the excitation process are not treated microscopically in the present effective model.

We next summarize the main perturbative conditions required for the coherent mapping. Coulomb interactions may shift the two-excitation manifold. Let $\Delta_C^{(\pm)}$ denote the Coulomb-induced shifts of the branches with $B$-channel OAM $\pm2$. The average shift $\bar{\Delta}_C=(\Delta_C^{(+)}+\Delta_C^{(-)})/2$ only renormalizes the two-photon resonance and can, in principle, be compensated by optical detuning. By contrast, the branch-dependent shift $\delta_C=\Delta_C^{(+)}-\Delta_C^{(-)}$ produces a relative phase. If $t_h$ is the characteristic time between creation of the two-excitation state and the heralding event, the corresponding overlap with the target state is
\begin{align}
F_C(t_h)
=
\cos^2
\left(
\frac{\delta_C t_h}{2\hbar}
\right)
\simeq
1-\frac{1}{4}
\left(
\frac{\delta_C t_h}{\hbar}
\right)^2 .
\label{eq:coulomb_branch_fidelity}
\end{align}
Thus, Coulomb-induced branch splitting is negligible when $|\delta_C|t_h/\hbar\ll1$.

Symmetry-breaking perturbations, such as confinement anisotropy, may also mix the two electronic OAM states. This effect can be represented by
\begin{align}
H_{\rm mix}
=
\lambda_C
\ket{+,+2}_{B}{}_{B}\bra{+,-2}
+
\lambda_C^{\ast}
\ket{+,-2}_{B}{}_{B}\bra{+,+2}.
\label{eq:oam_mixing_hamiltonian}
\end{align}
For short times, the unwanted transition probability scales as $(|\lambda_C|t_h/\hbar)^2$, giving
\begin{align}
1-F_{\rm mix}
=
O\left[
\left(
\frac{|\lambda_C|t_h}{\hbar}
\right)^2
\right].
\label{eq:oam_mixing_scaling}
\end{align}
In an ideally cylindrically symmetric disk, $|\psi_{+2}(\mathbf r)|^2=|\psi_{-2}(\mathbf r)|^2$, and angular-momentum conservation suppresses $\lambda_C$ to leading order. In this limit, $\delta_C$ and $\lambda_C$ vanish to leading order, while Coulomb interactions mainly produce a common two-excitation shift.

Finite-OAM excitations may also weakly leak radiatively into modes outside the selected output channel. We describe this loss phenomenologically by the annihilation of the corresponding finite-OAM excitation,
\begin{align}
L_{{\rm leak},\pm}
=
\sqrt{\Gamma_{\rm leak}}\,
X_{H,\pm2,B}.
\label{eq:finite_oam_leakage}
\end{align}
For $\Gamma_{\rm leak}t_h\ll1$, the probability that the stored OAM excitation is lost before heralding is approximately $\Gamma_{\rm leak}t_h$, and the ideal component is reduced as
\begin{align}
F_{\rm leak}(t_h)
\simeq
e^{-\Gamma_{\rm leak}t_h}
\simeq
1-\Gamma_{\rm leak}t_h .
\label{eq:finite_oam_leakage_fidelity}
\end{align}
The effective coherent mapping is valid in the perturbative regime
\begin{align}
\frac{|\delta_C|t_h}{\hbar}\ll1,\quad
\frac{|\lambda_C|t_h}{\hbar}\ll1,\quad
\Gamma_{\rm leak}t_h\ll1 .
\label{eq:validity_conditions_nonideal}
\end{align}

As a rough reference scale, two elementary charges separated by a lateral distance $r$ have $e^2/(4\pi\varepsilon_0\varepsilon_r r)\simeq0.1~{\rm meV}$ for GaAs with $\varepsilon_r\simeq12.9$ and $r\sim1~\mu{\rm m}$. The actual biexcitonic shift is determined by electron--hole envelope-function matrix elements and by cancellations between repulsive and attractive Coulomb terms, and thus need not coincide with this bare charge scale. A device-specific evaluation of $\bar{\Delta}_C$, $\delta_C$, $\lambda_C$, and $\Gamma_{\rm leak}$ using realistic OAM-mode profiles and optical mode functions is left for future microscopic modeling.

Radiative recombination of excitation $A$ maps its spin state back to the polarization of the emitted photon:
\begin{align}
    \ket{\downarrow,0}_{A} \rightarrow \ket{L,0}_{\gamma},\quad
    \ket{\uparrow,0}_{A} \rightarrow \ket{R,0}_{\gamma}.
    \label{eq:spin_to_photon_recombination}
\end{align}
The two-excitation state in Eq.~\eqref{eq:two_excitation_mapping} is then converted into
\begin{align}
\ket{\Psi_{\gamma B}}
=
\frac{1}{\sqrt{2}}
\left(
\ket{L,0}_{\gamma}\ket{+,+2}_{B}
+
\ket{R,0}_{\gamma}\ket{+,-2}_{B}
\right).
\label{eq:hybrid_state_with_spin}
\end{align}
Equation~\eqref{eq:hybrid_state_with_spin} is written in the full parallel notation. It shows that the photon polarization is correlated with the electronic OAM, while the emitted-photon OAM $l=0$ and the electronic spin state $\ket{+}_{s,B}$ are common to both branches. Thus, the entangled degrees of freedom are the photon polarization and the electronic OAM.

\begin{figure}[t]
\includegraphics[width=1\linewidth]{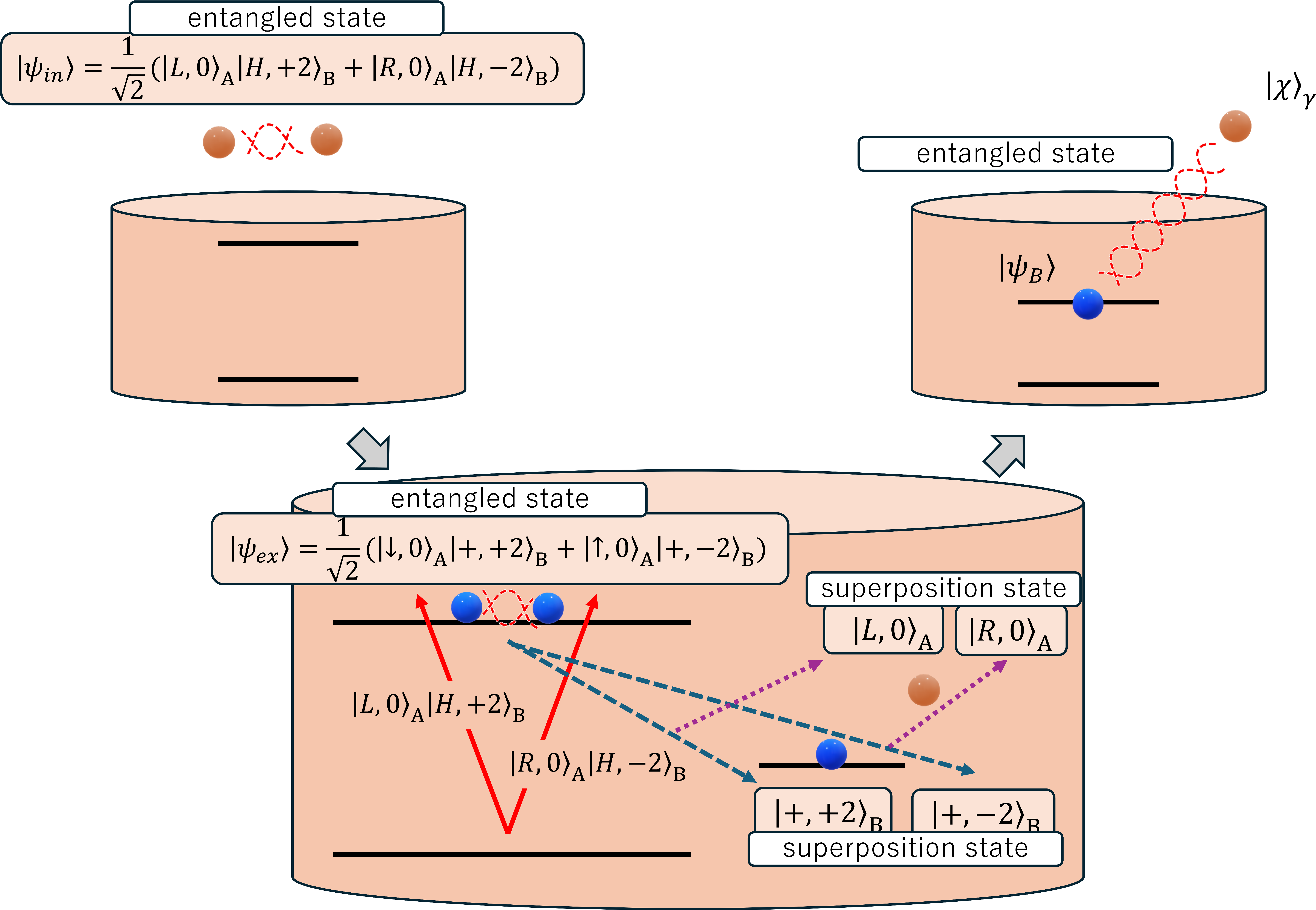}
\caption{Conceptual illustration of photon--electron hybrid-entanglement generation within a single semiconductor quantum disk. The labels $A$ and $B$ denote two optically addressed excitation channels in the same disk: a zero-OAM channel that recombines radiatively and a finite-OAM channel that stores the residual electronic OAM qubit. A spin--orbit structured two-photon state is mapped to a two-excitation state through optical spin and OAM selection rules. Subsequent radiative recombination of the zero-OAM excitation emits a photon whose polarization is entangled with the residual electronic OAM state.}
\label{fig:hybrid_entanglement_scheme}
\end{figure}

Figure~\ref{fig:hybrid_entanglement_scheme} summarizes the effective coherent mapping in Eq.~\eqref{eq:two_excitation_mapping} and the recombination process in Eq.~\eqref{eq:spin_to_photon_recombination}. It is useful to distinguish the photon--electron entangled state from the electronic state obtained after measuring the emitted photon. If the photon is projected onto
\begin{align}
    \ket{\chi}_{\gamma}
    =
    \alpha \ket{L,0}_{\gamma}
    \pm
    \beta \ket{R,0}_{\gamma},
    \quad
    |\alpha|^2+|\beta|^2=1,
    \label{eq:photon_projection_state}
\end{align}
then the residual electronic state becomes
\begin{align}
    \ket{\psi_B}
    =
    \alpha^{\ast}\ket{+,+2}_{B}
    \pm
    \beta^{\ast}\ket{+,-2}_{B},
\label{eq:conditional_electron_state}
\end{align}
up to normalization and an overall phase. Since the spin state $\ket{+}_{s,B}$ is common to both terms, Eq.~\eqref{eq:conditional_electron_state} represents the conditional electronic OAM qubit with a factored-out spin component. Thus, Eq.~\eqref{eq:hybrid_state_with_spin} describes the photon--electron entangled state before photon measurement, whereas Eq.~\eqref{eq:conditional_electron_state} describes the electronic state conditioned on the photon-polarization measurement in Eq.~\eqref{eq:photon_projection_state}.

The effective-state mappings above summarize the basic mechanism of the proposed scheme. Additional material- and geometry-dependent processes, such as pure OAM dephasing, phonon-assisted relaxation, readout imperfections, and leakage to other dark or bright states, may further reduce the fidelity in a device-dependent manner. These effects are outside the present minimal model and should be included in future device-specific studies. We next analyze the corresponding open-system dynamics using a master-equation approach and evaluate the heralded fidelity of the generated photon--electron entangled state.

\section{Simulation}
\label{sec:simulation}

\begin{figure}[t]
\includegraphics[width=\linewidth]{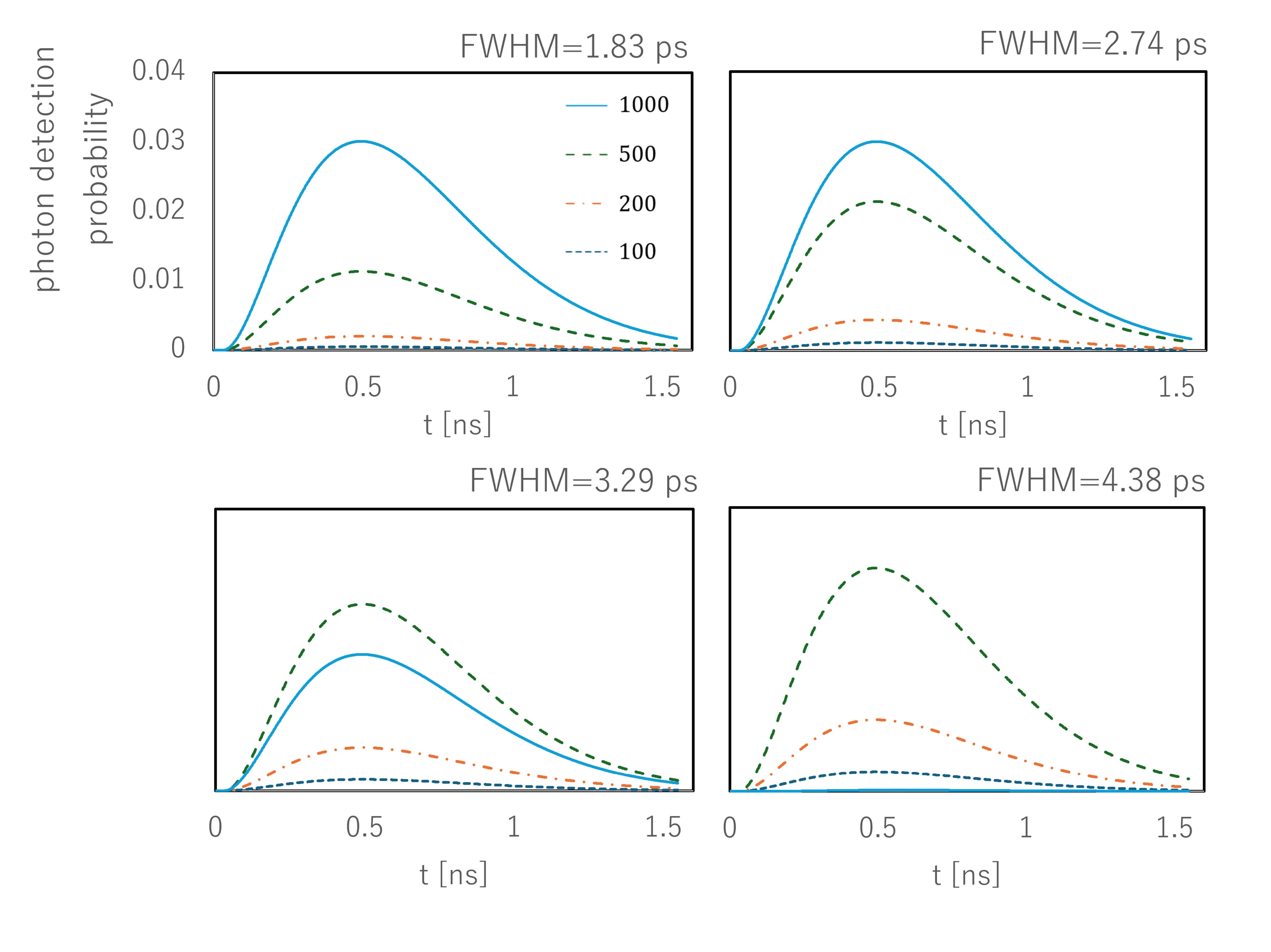}
\caption{Time evolution of the single-photon probability in the selected output filter mode for incident pulses with different full widths at half maximum (FWHM). Results are shown for $g/\Gamma_{\rm rad}=$100 (blue dash-dot-dot), 200 (orange dash-dotted), 500 (green dash), and 1000 (light-blue solid). The plotted quantity is the single-photon occupation probability within the filter-mode description and should be interpreted together with the conditional fidelity. External collection losses, propagation losses, mode-matching losses, and detector inefficiencies are not included.}
\label{fig:single_photon_probability}
\end{figure}

We analyze the open-system dynamics of the proposed hybrid-entanglement generation scheme using a quantum master-equation approach. The purpose of the simulation is to evaluate the emission dynamics, the heralded fidelity of the generated photon--electron entangled state, and the effect of orbital relaxation before the heralding event. Consistent with the effective-model framework introduced in Sec.~\ref{sec:model}, the incident spin--orbit structured two-photon pulse is represented as a time-dependent drive from the zero-excitation state to the selected two-excitation state. The microscopic two-photon absorption process is not treated explicitly; instead, the simulation starts from an effective drive to the selected coherent two-excitation channel.

We assume that radiative recombination of the zero-OAM excitation in channel $A$ dominates over recombination processes involving the finite-OAM excitation in channel $B$. Thus, on the emission timescale considered here, only excitation $A$ emits a photon into the selected output mode, while the electronic orbital degree of freedom in channel $B$ remains as the stationary qubit. This assumption corresponds to the ideal mode-selective limit introduced in Eq.~\eqref{eq:recombination_example}. Finite-OAM radiative leakage and other material-dependent decay channels are not included in the numerical simulation; their perturbative effects were estimated in Sec.~\ref{sec:model}, except for the orbital-relaxation channel introduced below.

For channel $A$, the optically active excitations $\ket{\uparrow,0}_A$ and $\ket{\downarrow,0}_A$ emit right- and left-circularly polarized photons, respectively. The corresponding radiative decay channels are described by the jump operators
\begin{align}
L_R
=
\sqrt{\Gamma_{\rm rad}}\,X_{R,0,A},
\quad
L_L
=
\sqrt{\Gamma_{\rm rad}}\,X_{L,0,A}.
\label{eq:radiative_jump_operators}
\end{align}
For simplicity, we take the two radiative decay rates to be equal. The value of $\Gamma_{\rm rad}$ is estimated from the inverse of the exciton lifetime, $\tau=1.55$ ns~\cite{PhysRevB.72.035314}.

To describe the photon emitted into the selected output channel, we introduce an auxiliary output filter mode $\gamma$. This mode represents the detected spatial or fiber mode rather than the full continuum of radiation modes. For each circular polarization $\mu=R,L$, the filter mode is represented by a bosonic annihilation operator $b_\mu$. The free Hamiltonian of the filter mode is
\begin{align}
H_f=\sum_{\mu=R,L}\omega_f b_\mu^\dagger b_\mu .
\label{eq:filter_hamiltonian}
\end{align}
In the numerical calculation, we work in a rotating frame resonant with the optical transition, so that the relevant detuning is absorbed into the definition of $\omega_f$.

The incident spin--orbit structured two-photon pulse is not modeled as a microscopic sequence of single-photon absorption events. Instead, after eliminating the optical preparation stage and possible intermediate virtual states, we represent it by a time-dependent effective drive from the zero-excitation state to the selected coherent two-excitation state,
\begin{align}
H_p(t)=g f(t)
\left(
\ket{\psi_{\rm ex}}\bra{G}
+
\ket{G}\bra{\psi_{\rm ex}}
\right),
\label{eq:pulse_hamiltonian}
\end{align}
where $\ket{\psi_{\rm ex}}$ is the normalized two-excitation state corresponding to the ideal coherent mapping in Eq.~\eqref{eq:two_excitation_mapping}. The parameter $g$ is an effective two-photon Rabi coupling to this selected collective channel. This single-coupling description is valid when the two absorption branches in Eq.~\eqref{eq:two_excitation_mapping} have equal effective two-photon amplitudes up to a fixed calibratable phase, when the absorption process does not leave which-path information in intermediate or environmental degrees of freedom, and when branch-dependent detunings, OAM mixing, and leakage are negligible on the pulse and heralding timescales. If these conditions are not satisfied, Eq.~\eqref{eq:pulse_hamiltonian} should be replaced by a branch-dependent drive with distinct couplings, phases, and detunings.

The Gaussian pulse envelope is
\begin{align}
f(t)=\exp\left[-\frac{(t-t_0)^2}{2\sigma^2}\right],
\label{eq:gaussian_pulse}
\end{align}
where the full width at half maximum is
\begin{align}
{\rm FWHM}=2\sqrt{2\ln 2}\,\sigma .
\label{eq:fwhm_definition}
\end{align}

The unidirectional coupling from the radiative emission channel of channel $A$ to the selected output filter mode is described using a cascaded-systems master equation~\cite{gardiner1993driving,carmichael1993quantum}. Here, $H_{AB}$ denotes the effective Hamiltonian of the matter degrees of freedom in channels $A$ and $B$ in the rotating frame. Branch-independent energies are absorbed into the rotating frame. The time evolution of the total density matrix $\rho(t)$ is given by
\begin{align}
\dot{\rho}(t)
=&-i[H_{AB}+H_p(t)+H_f+H_{\rm cas},\rho]
\nonumber\\
&+\sum_{\mu=R,L}
\left(
\mathscr{D}[L_\mu]\rho
+
\mathscr{D}[\sqrt{\kappa}b_\mu]\rho
\right.
\nonumber\\
&\left.
+
[\sqrt{\kappa}b_\mu\rho,L_\mu^\dagger]
+
[L_\mu,\rho\sqrt{\kappa}b_\mu^\dagger]
\right),
\label{eq:master_equation}
\end{align}
where
\begin{align}
H_{\rm cas}
=
\frac{i}{2}
\sum_{\mu=R,L}
\left(
L_\mu^\dagger\sqrt{\kappa}b_\mu
-
\sqrt{\kappa}b_\mu^\dagger L_\mu
\right)
\label{eq:cascaded_hamiltonian}
\end{align}
is the coherent part of the cascaded coupling. The jump operators $L_\mu$ describe radiative emission from channel $A$, while $\sqrt{\kappa}b_\mu$ describes leakage from the auxiliary filter mode. Thus, the filter mode represents the selected output mode fed by the radiative channel, rather than an independent emitter. Here, $\kappa$ is the decay rate of the filter mode, and
\(
\mathscr{D}[O]\rho
=
O\rho O^\dagger
-
\frac{1}{2}
\left\{
O^\dagger O,\rho
\right\}
\)
denotes the Lindblad dissipator. Throughout the calculation we set $\kappa=10\Gamma_{\rm rad}$, corresponding to a filter bandwidth larger than the radiative linewidth. We confirmed that moderate changes in $\kappa$ do not qualitatively affect the heralded fidelity.

To obtain the conditional state of the emitted photon and the residual electronic orbital degree of freedom, we trace out channel $A$:
\begin{align}
\rho_{B\gamma}(t)={\rm Tr}_A[\rho(t)].
\label{eq:reduced_density_matrix}
\end{align}
We then project the filter mode onto the single-photon subspace,
\begin{align}
\Pi_1
=
\ket{1_R,0_L}\bra{1_R,0_L}
+
\ket{0_R,1_L}\bra{0_R,1_L},
\label{eq:single_photon_projector}
\end{align}
and define the normalized conditional state
\begin{align}
\tilde{\rho}_{B\gamma}(t)
=
\frac{
(\bm{I}_B\otimes\Pi_1)
\rho_{B\gamma}(t)
(\bm{I}_B\otimes\Pi_1)
}{
P_{1{\rm ph}}(t)
},
\label{eq:conditional_density_matrix}
\end{align}
where
\begin{align}
P_{1{\rm ph}}(t)
=
{\rm Tr}
\left[
(\bm{I}_B\otimes\Pi_1)\rho_{B\gamma}(t)
\right]
\label{eq:single_photon_probability}
\end{align}
is the single-photon probability in the selected output filter mode. Thus, $\tilde{\rho}_{B\gamma}(t)$ describes the state conditioned on the presence of one photon in the selected output mode at time $t$, whereas $P_{1{\rm ph}}(t)$ characterizes the corresponding single-photon probability within the filter-mode description. The fidelity defined below is a conditional, heralded fidelity and should not be interpreted as the unconditional fidelity of the full system. The plotted single-photon probability does not include external collection losses, propagation losses, mode-matching losses outside the selected filter mode, or detector inefficiencies; these factors would reduce the experimentally observed heralding rate.

The target entangled state is defined as
\begin{align}
\ket{\Psi_{\rm ent}}
=
\frac{1}{\sqrt{2}}
\left(
\ket{L,0}_{\gamma}\ket{+,+2}_{B}
+
\ket{R,0}_{\gamma}\ket{+,-2}_{B}
\right).
\label{eq:target_bell_state}
\end{align}
In the numerical evaluation, this target vector is represented in the reduced polarization--electronic-OAM subspace by suppressing the common emitted-photon OAM $l=0$ and the common spin state $\ket{+}_{s,B}$. The heralded entanglement fidelity is then evaluated as
\begin{align}
F(t)
=
\bra{\Psi_{\rm ent}}
\tilde{\rho}_{B\gamma}(t)
\ket{\Psi_{\rm ent}}.
\label{eq:heralded_fidelity}
\end{align}
This quantity characterizes the photon--electron entangled state at the heralding time, when a single photon is detected in the selected output mode. It is distinct from the subsequent storage fidelity of the electronic OAM qubit after photon detection. Orbital relaxation after the heralding event corresponds to memory decay and is not included in $F(t)$; by contrast, orbital relaxation before heralding reduces the conditional fidelity, as analyzed below. At very early times, where $P_{1{\rm ph}}(t)$ is negligibly small, the normalized conditional state is not physically relevant. In the numerical evaluation of $F(t)$, times at which $P_{1{\rm ph}}(t)$ is below a small numerical threshold are excluded.

To illustrate the dynamics within this effective description, we examine representative coupling strengths $g/\Gamma_{\rm rad}=100$--$1000$ and several pulse widths. The experimental Rabi frequency reported in Ref.~\cite{terashima2025optical} is used only as a reference scale for coherent optical manipulation in semiconductor quantum disks; the effective two-photon coupling $g$ in Eq.~\eqref{eq:pulse_hamiltonian} is not identified directly with a microscopic single-photon Rabi frequency. Its value is device- and pulse-configuration-dependent and should ultimately be obtained from the relevant two-photon matrix element, intermediate-state detunings, and optical-mode overlap integrals.

\begin{figure}[t]
\includegraphics[width=\linewidth]{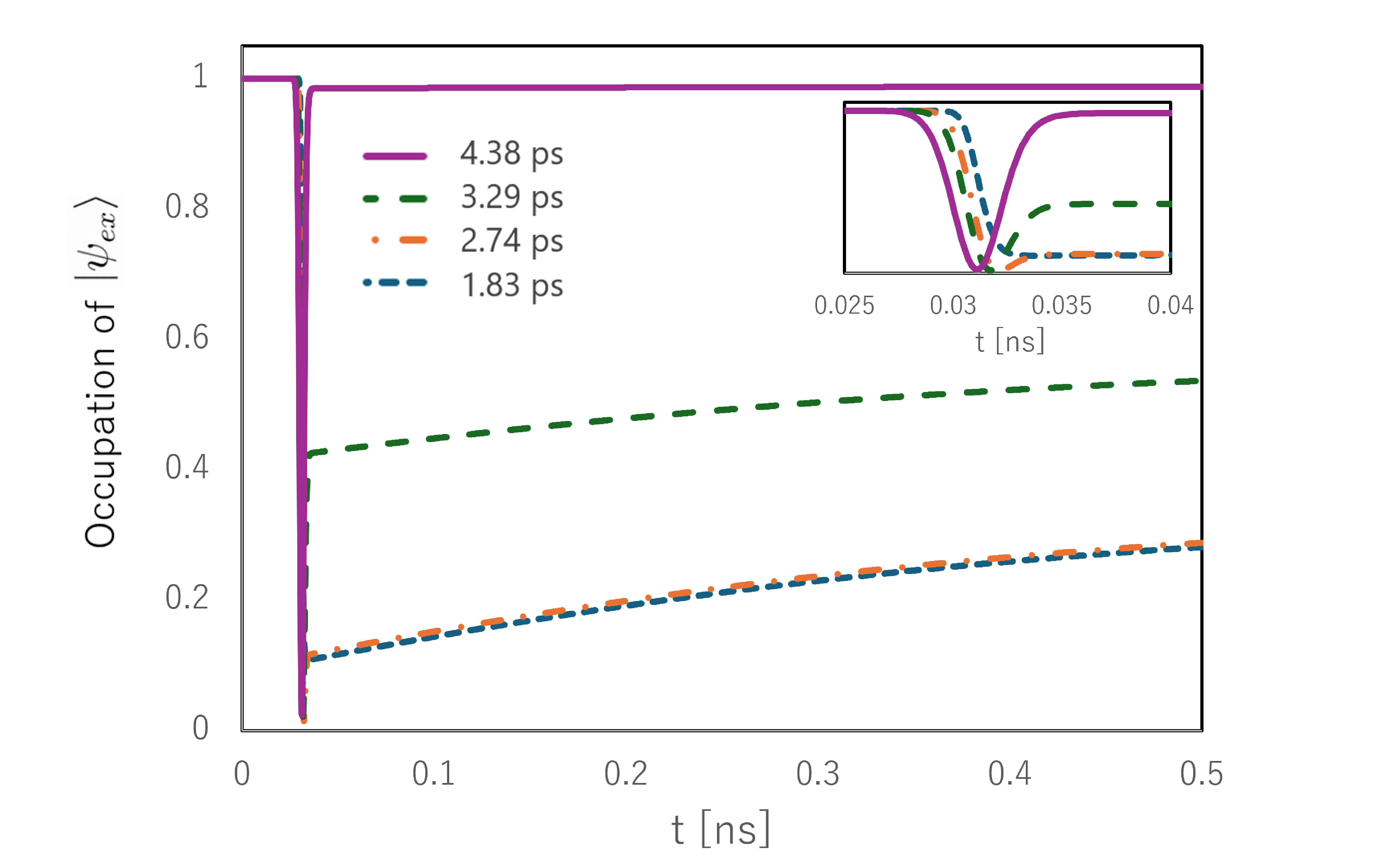}
\caption{Time evolution of the occupation probability of the selected two-excitation state $\ket{\psi_{\rm ex}}$ for different incident pulse widths at $g/\Gamma_{\rm rad}=1000$. The purple solid, green dashed, orange dash-dotted, and blue dotted lines correspond to FWHM=4.38 ps, 3.29 ps, 2.74 ps, and 1.83 ps, respectively. Inset: Enlarged view of the region where the dynamics change rapidly ($t=0.025$--$0.040$ ns).}
\label{fig:two_excitation_occupation}
\end{figure}

Figure~\ref{fig:single_photon_probability} shows the single-photon probability $P_{1{\rm ph}}(t)$ in the selected output filter mode. The result depends strongly on both $g$ and the pulse width, reflecting the pulse-area dependence of the effective excitation process. Changing either parameter modifies the occupation of the selected two-excitation channel and leads to Rabi-like oscillations in the photon probability. Figure~\ref{fig:two_excitation_occupation} shows the corresponding occupation probability of the selected two-excitation state $\ket{\psi_{\rm ex}}$ for $g/\Gamma_{\rm rad}=1000$, confirming the same pulse-area dependence before radiative recombination. The rapid initial change reflects coherent transfer between $\ket{G}$ and $\ket{\psi_{\rm ex}}$, while the subsequent revival is associated with back-transfer within the effective two-level subspace.

To examine the effect of orbital relaxation before the heralding event, we introduce additional Lindblad operators describing nonradiative relaxation from the finite-OAM states to the electronic zero-OAM state in channel $B$,
\begin{align}
L_{{\rm orb},\pm}
=
\sqrt{\Gamma_{\rm orb}}\,
\ket{+,0}_{B}{}_{B}\bra{+,\pm 2}.
\label{eq:orbital_relaxation_operators}
\end{align}
Here, $\Gamma_{\rm orb}$ denotes the orbital relaxation rate. The channel in Eq.~\eqref{eq:orbital_relaxation_operators} is included as a representative mechanism that directly erases the OAM information stored in channel $B$ while leaving the spin state unchanged. Other device-dependent imperfections, including OAM dephasing, phonon-assisted dephasing, readout backaction, and leakage channels beyond those considered in Sec.~\ref{sec:model}, are not included in the minimal simulation. These mechanisms may further reduce the fidelity in a device-dependent manner, but do not alter the basic entanglement-generation mechanism.

\begin{figure}[t]
\centering
\includegraphics[width=0.9\linewidth]{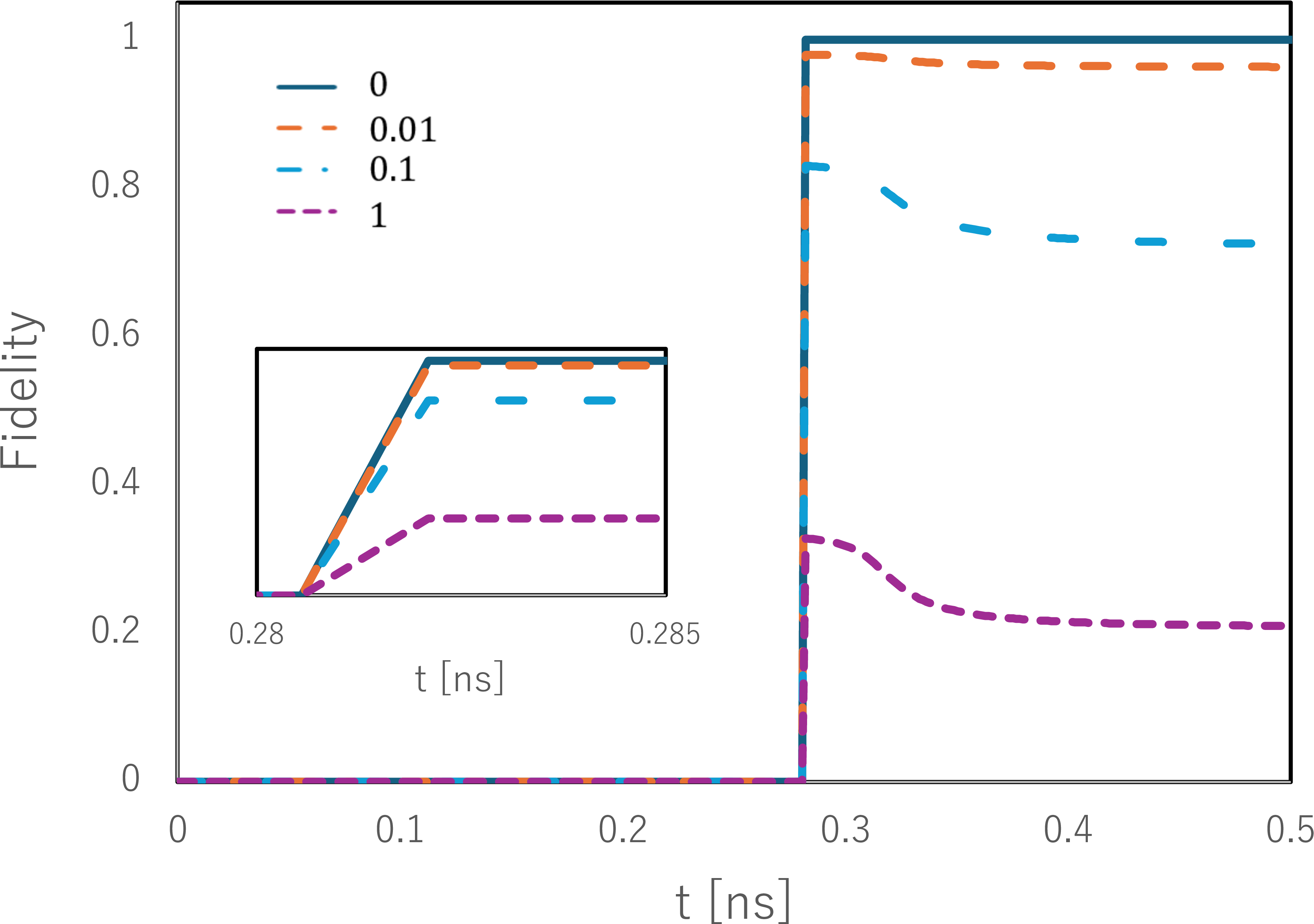}
\caption{Time evolution of the heralded entanglement fidelity for different orbital relaxation rates $\Gamma_{\rm orb}$. The blue solid, orange dashed, light-blue dash-dotted, and purple dotted lines correspond to $\Gamma_{\rm orb}/\Gamma_{\rm rad}$=0, 0.01, 0.1, and 1, respectively. The fidelity is evaluated after projection onto the single-photon subspace of the selected output mode. The data shown correspond to $g/\Gamma_{\rm rad}=1000$ and FWHM = 1.83 ps. The fidelity should be interpreted only in the time window where the single-photon probability is appreciable. Orbital relaxation before the heralding event degrades the conditional state by erasing the OAM information in channel $B$. Inset: Enlarged view of the region where the dynamics change rapidly ($t =$ 0.2--0.4 ns).}
\label{fig:fidelity_relaxation}
\end{figure}

Figure~\ref{fig:fidelity_relaxation} shows the heralded entanglement fidelity for different orbital relaxation rates. In the ideal case without orbital relaxation, $\Gamma_{\rm orb}=0$, the conditional state reaches nearly unit fidelity with the target entangled state in Eq.~\eqref{eq:target_bell_state} once a photon is present in the selected output mode. As $\Gamma_{\rm orb}$ increases, relaxation from $\ket{+,+2}_B$ or $\ket{+,-2}_B$ to $\ket{+,0}_B$ before the heralding event removes the stored orbital information and degrades the photon--electron OAM correlation. High-fidelity heralded hybrid entanglement requires the orbital relaxation time to be longer than the characteristic emission and detection timescale. Once the photon has been detected, subsequent relaxation of the electronic OAM state should be regarded as memory decay after heralding, which is separate from the heralded fidelity evaluated here. The abrupt rise in the plotted fidelity reflects the normalization of the state after projection onto the single-photon subspace; therefore, the fidelity should be interpreted together with $P_{1{\rm ph}}(t)$.

\section{Conclusion}
\label{sec:conclusion}

We proposed a minimal quantum-optical scheme for generating hybrid entanglement between photon polarization and electronic orbital angular momentum in a semiconductor quantum disk. The scheme uses two optically addressed excitation channels within the same disk: a zero-OAM channel that recombines radiatively and a finite-OAM channel that stores the residual electronic orbital qubit. A spin--orbit structured two-photon state is mapped to a selected two-excitation state through optical selection rules. Subsequent radiative recombination emits a photon whose polarization is entangled with the remaining electronic orbital state. In the ideal coherent limit, the heralded state conditioned on single-photon occupation of the selected output mode approaches the target hybrid entangled state.

Using a master-equation model, we analyzed the emission dynamics, the selected-mode single-photon probability, and the heralded entanglement fidelity. The single-photon probability is governed by the effective coupling strength and pulse duration through the pulse-area dependence of the excitation process. Orbital relaxation before heralding degrades the fidelity by erasing the stored orbital information, indicating that high-fidelity operation requires the orbital relaxation time to exceed the relevant emission and detection timescale. We also discussed perturbative validity conditions for branch-dependent Coulomb shifts, symmetry-breaking OAM mixing, and finite-OAM radiative leakage. These effects remain small when $|\delta_C|t_h/\hbar\ll1$, $|\lambda_C|t_h/\hbar\ll1$, and $\Gamma_{\rm leak}t_h\ll1$, respectively.

The proposed mechanism converts the spin--orbit structure of light into photon--electron hybrid entanglement through radiative recombination, without requiring direct measurement of the electronic state during the generation step. The emitted photon serves as a flying qubit, while the residual finite-OAM electronic excitation provides a stationary orbital qubit in the semiconductor nanostructure. The present work should be understood as a proof-of-principle effective model. Further progress will require device-specific microscopic modeling of the two-photon matrix element, realistic collection and detection efficiencies, orbital-state lifetimes and readout schemes, and possible extensions to remote-entanglement protocols based on two-photon interference and entanglement swapping~\cite{park2020entanglement,tsujimoto2018high}. These developments would clarify the conditions under which structured-light-mediated photon--electron hybrid entanglement can be implemented in semiconductor platforms.

\section*{Acknowledgments}

This work was supported by the JSPS KAKENHI Grants No.~JP22K04863, No.~JP22H05131 and No.~JP22H05132, and by the JSPS International Joint Research Program JRP- LEAD with UKRI under Grant No.~JPJSJRP20241710.

\bibliography{hoge}

\end{document}